\documentclass{elsart}

\usepackage{epsfig}
\usepackage[english]{babel}
\usepackage{amssymb}

\begin{document}

\begin{frontmatter}

\title{A PMT-Block Test Bench}

\author[label1,label7,label10]{P. Adragna},
\author[label2,label3]{A. Antonaki},
\author[label4]{I. Boudagov},
\author[label1]{V. Cavasinni},
\author[label1,label8]{D. Costanzo},
\author[label1]{T. Del Prete},
\author[label1]{A. Dotti},
\author[label3]{D. Fassouliotis},
\author[label2,label3]{V. Giakoumopoulou},
\author[label2,label3]{N. Giokaris},
\author[label6]{C. Guicheney},
\author[label2,label3]{E. Karali},
\author[label4,label5]{J. Khubua},
\author[label2,label3]{M. Lebessi},
\author[label1]{A. Lupi},
\author[label3]{A. Manoussakis},
\author[label1]{E. Mazzoni},
\author[label4,label5]{I. Minashvili},
\author[label2,label3]{M. Morphi},
\author[label1]{G.F. Pagani},
\author[label1]{R. Paoletti},
\author[label1]{D. Rizzi},
\author[label1]{C. Roda},
\author[label1]{F. Sarri},
\author[label2,label3]{Ath. Staveris-Polykalas},
\author[label2,label3]{Th. Staveris-Polykalas},
\author[label4]{S. Stoudenov},
\author[label1,label9]{G. Usai},
\author[label6]{F. Vazeille},
\author[label2,label3]{C. Vellidis},
\author[label2,label3]{I. Vichou},
\author[label1]{I. Vivarelli},
\author[label1]{M. Volpi},

\address[label1]{Dipartimento di Fisica "E.Fermi", Universita di Pisa and 
  Istituto Nazionale di Fisica Nucleare, Sezione di Pisa, Largo B. Pontecorvo
  3, Pisa 56127, Italy}
\address[label3]{National Capodistrian University of Athens, 
  30 Panepistimiou st., Athens 10679, Greece}
\address[label2]{Institute of Accelerating Systems and Applications, 
  P.O. Box 17214, Athens 10024, Greece}
\address[label4]{Joint Institute for Nuclear Research, Dubna, 
Moscow Region, 141980, Russia}
\address[label5]{Institute of High Energy Physics of Tbilisi University, 
  9 University st., 380086, Tbilisi, Georgia}
\address[label6]{LPC Clermont-Ferrand, Universite' Blaise Pascal/CRNS-IN2P3,
Clermont-Ferrand, France}
\address[label7]{Universita' degli studi di Siena, via Roma 56, 53100 Siena,
  Italia.}
\address[label10]{Now with Queen Mary,
  University of London, Mile End Road London, E1 4NS, UK}
\address[label8]{University of Sheffield, 
  Hounsfield Rd, Sheffield, S3 7RH, UK}
\address[label9]{Now with University of Chicago, Enrico Fermi Institute, 
  5640 South Ellis Avenue, Chicago, Illinois 60637, USA}

\begin{abstract}
The front-end electronics of the ATLAS hadronic calorimeter (Tile Cal)  
is housed in a unit, called {\it PMT-Block}.
The PMT-Block is a compact instrument comprising a light mixer, 
a PMT together with its divider and a {\it 3-in-1} card, 
which provides shaping, amplification and
integration  for the  signals.
This instrument needs to be qualified before being assembled on the detector.
A PMT-Block test bench has been developed for this purpose. 
This test bench is a system which allows fast, albeit accurate enough,
measurements of the main properties of a complete PMT-Block.
The system, both hardware and software, 
and the protocol used for the PMT-Blocks characterisation are 
described in detail in this report.
The results obtained in the test 
of about 10000 PMT-Blocks needed for the instrumentation of 
the ATLAS (LHC-CERN) hadronic Tile Calorimeter are also reported. 
 
\end{abstract}

\end{frontmatter}

\section{Introduction}
\label{sec:introduction}
A PMT-Block \cite{bib:TDR} \cite{bib:PMTB} 
is the device used by the ATLAS Tile Calorimeter 
to convert light into electric signals. 
It is composed of a photomultiplier tube\footnote{\it Hamamatsu 7877}, 
a light mixer,
a high voltage divider and a {\it 3-in-1} card \cite{bib:treinuno}.
The light mixer provides the interface between the PMT and the fiber 
bundle which collects some of
the light produced in the scintillator of the sampling calorimeter. 
The {\it 3-in-1} card forms the front-end of the electronic
read-out chain. 
It provides three basic functions: fast pulse shaping and amplification to 
accommodate the large dynamic range needed by the detector, 
charge injection calibration and
slow integration of the PMT signals for monitoring and calibrations.
This paper will not enter into a detailed description of each element. 
A documentation
of their characteristics has been already published for most of them.

\begin{figure}[ht]
  \begin{minipage}{.495\linewidth}
    \begin{center}
      \epsfig{file=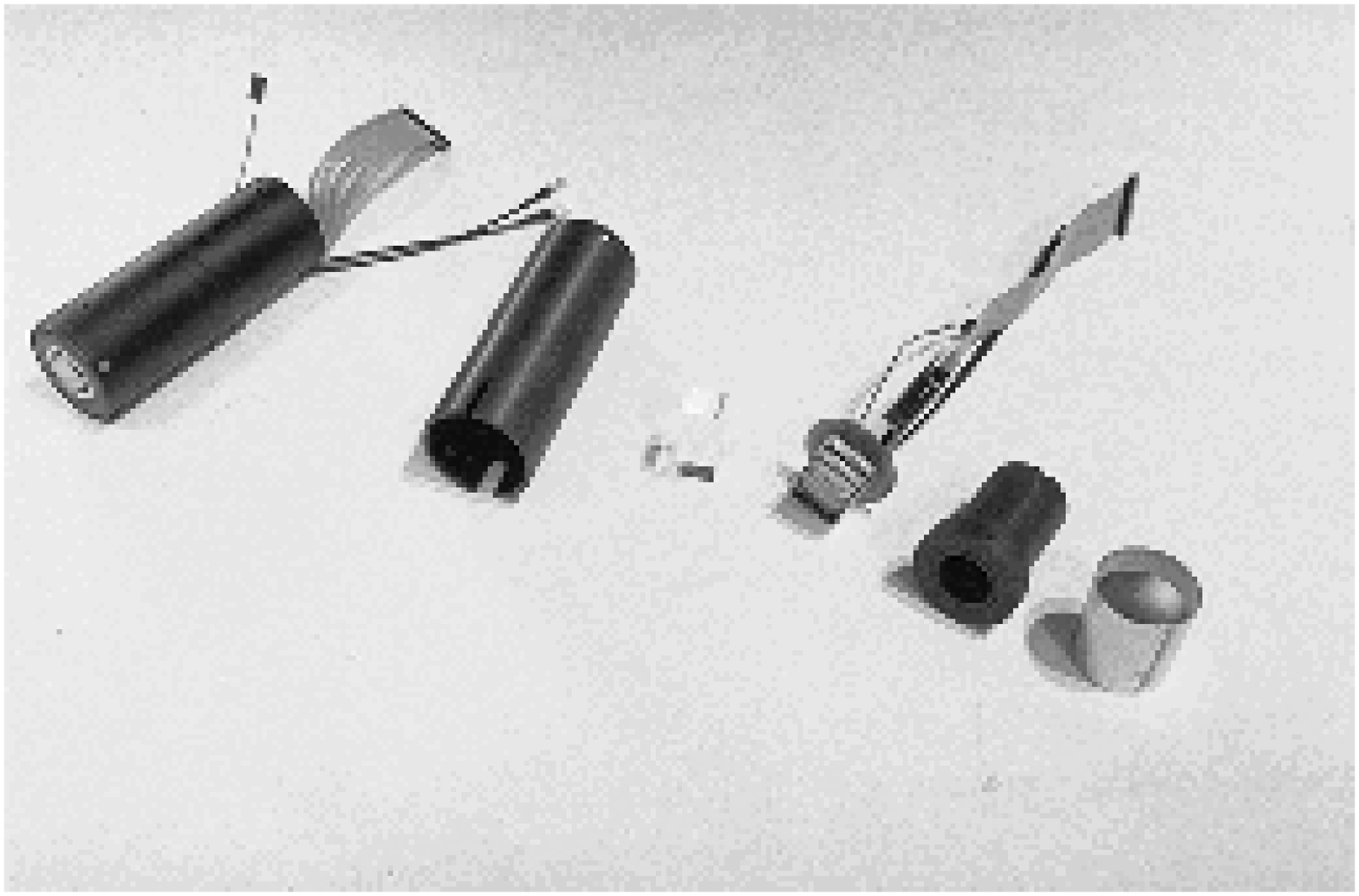,width=0.5\linewidth,angle=-90}
    \end{center}
  \end{minipage}
  \begin{minipage}{.495\linewidth}
    \begin{center}
      \epsfig{file=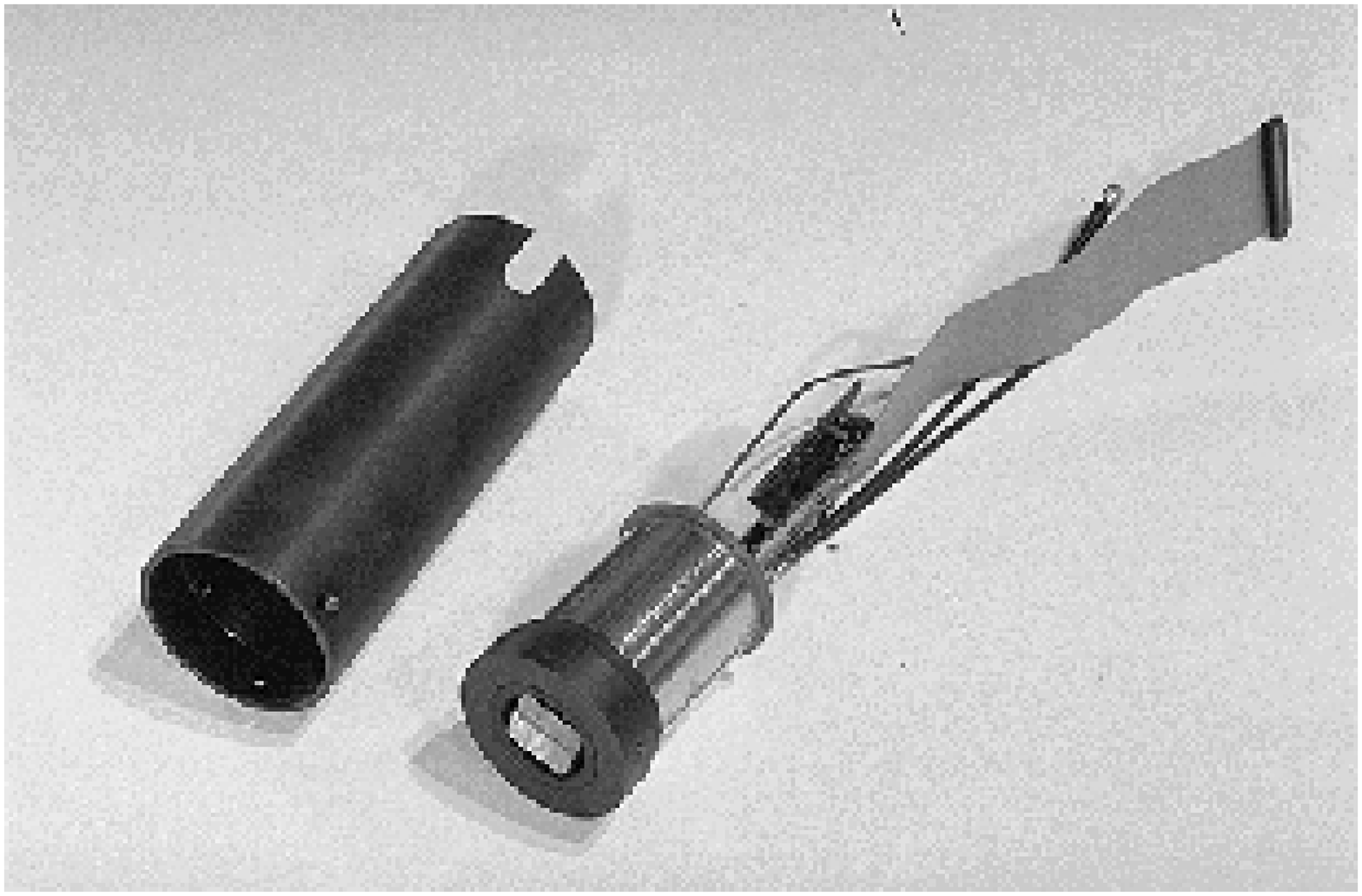,width=0.5\linewidth,angle=-90}
    \end{center}
  \end{minipage}
  \caption{\sl\small The PMT-Block. a) on the left the individual pieces
    are shown and b) on the right the PMT-Block is fully assembled.}
  \protect\label{fig:pmtblock}
\end{figure}

A PMT is assembled inside a soft iron and a mu-metal cylinder, 
which provides adequate magnetic shielding. 
The individual components and 
the fully assembled PMT-Block are shown in
figure \ref{fig:pmtblock}.

The PMT-Blocks are housed inside {\it drawers}.
A drawer provides the mechanical support both for the PMT-Block and 
the electronic boards (also called {\it mother boards}).
A drawer contains up to 24 PMT-Blocks.
The mother boards are
connected to each PMT-Block, supplying the PMT dividers
with the high voltage,
the {\it 3-in-1} cards with the low voltages and conveying 
control and calibration signals to the {\it 3-in-1} cards. 
The mother board also houses the electronics to process the signals coming
from the shaper and the integrator.

All the components assembled in a PMT-Block underwent an individual 
Quality Control process. However, it is necessary to check 
the PMT-Block as a whole before inserting it in 
a drawer. 
This test bench is intended to check each specific functionality 
of a PMT-Block while no attempt is made to characterise the 
individual components, as specific test benches are dedicated to this task. 
The result of the test here described is either a certification of 
full functionality, or the 
identification of the parts of the PMT-Block 
which are not working properly, together with a 
brief description of the problem. 

The PMT-Block test bench has been developed to be: 
\begin{description}
\item[--] {\it Affordable}. 
  Commercial electronics can be very expensive and should be 
  avoided whenever it is possible. On the other side spare electronics 
  is available in the laboratories and can be employed for the tests.

\item[--] {\it Easy to use}. The system should be operated also by a 
non-expert crew, hence it must be very intuitive and easy to use. 

\item[--] {\it Stable}. A few years of work were needed to characterize 
  all the  PMT-Blocks. 
  During this time, the system had to operate in a stable
  way, with no major breakdowns. 
  Moreover, the test bench will be kept working during all 
  of the experiment's lifetime for test purposes.
  This implies that the test bench must operate in stable conditions 
  for about 15 years.

\item[--] {\it Portable}. 
  Since the test bench has to operate for many years, we expect
  that the platform used for the DAQ system will change. It is desirable that
  the high level functionality of the test bench will not change, leaving the
  low level to cope with a changing platform. 
  It has been decided to write the DAQ in ANSI C code to minimise 
  the efforts to move the system to another platform and to reduce 
  the costs that another institution would bear if a commercial 
  DAQ system was used. 

\item[--] {\it Versatile}. 
  The DAQ system has to be easily modifiable and versatile 
  enough to accomplish any change in the protocol and also in the hardware 
  itself. 
  This characteristic turned out to be very useful, since the same DAQ 
  has been used also in other test benches for PMT characterization. 
\end{description}

Keeping this in mind, the design of the test bench was split 
into three parts:  
\begin{description}
\item[--] The hardware used in this test bench 
\item[--] The protocol used for the characterization of the PMT-Block test  
\item[--] The DAQ software which is rather independent 
  of this specific test bench 
\end{description}

In this paper each one item of the above items is described in detail 
and the tests performed are  
reported. 
\section{The experimental set-up}
\label{sec:setup}
\begin{figure}[ht]
  \begin{center}
    \epsfig{file=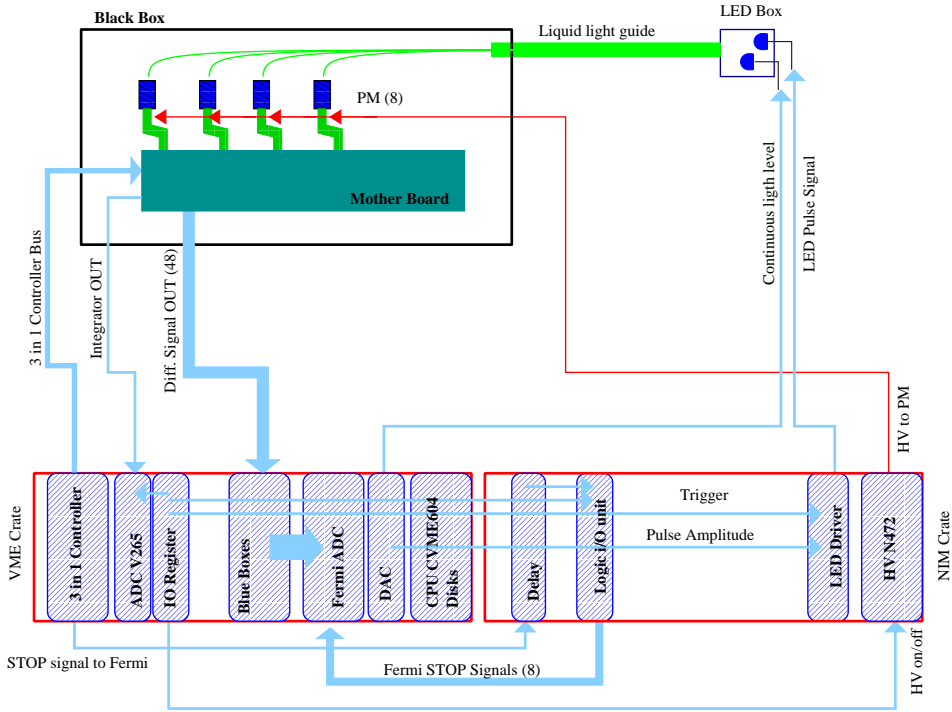,width=0.9\linewidth,angle=0}
  \end{center}
  \caption{\sl\small Schematic diagram of the test bench.}
  \protect\label{fig:testbench}
\end{figure}

\begin{figure}[ht]
  \begin{center}
    \epsfig{file=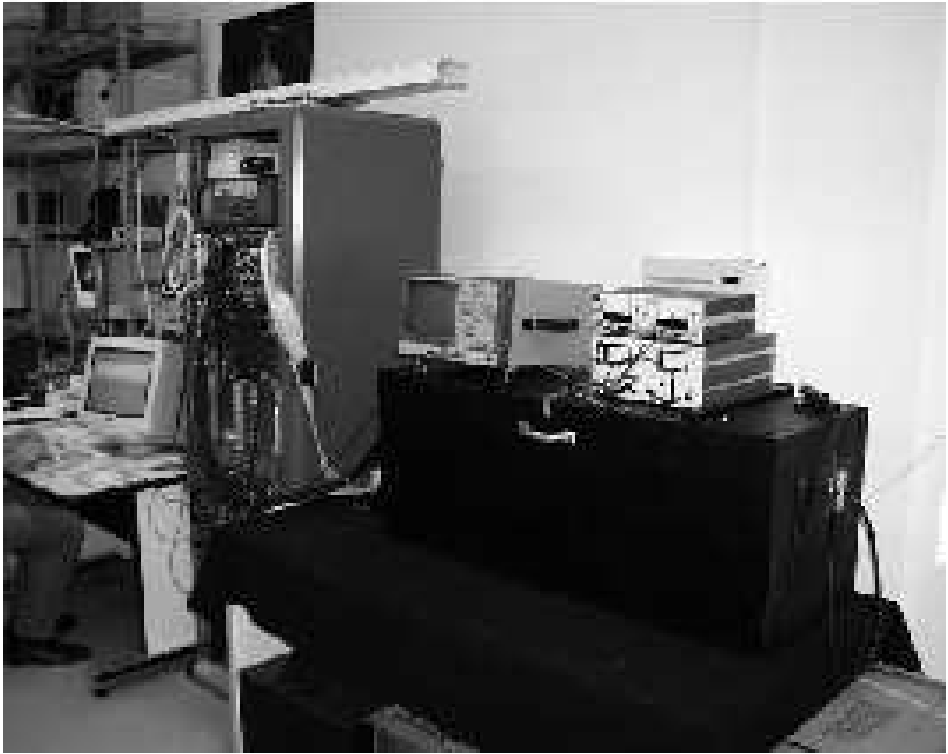,width=0.9\linewidth,angle=0}
  \end{center}
  \caption{\sl\small Overall view of the test bench. The black
    box, closed, on the right. On the left the electronics for the DAQ.}
  \protect\label{fig:blackbox}
\end{figure}

\begin{figure}[ht]
  \begin{minipage}{0.495\linewidth}
    \begin{center}
      \epsfig{file=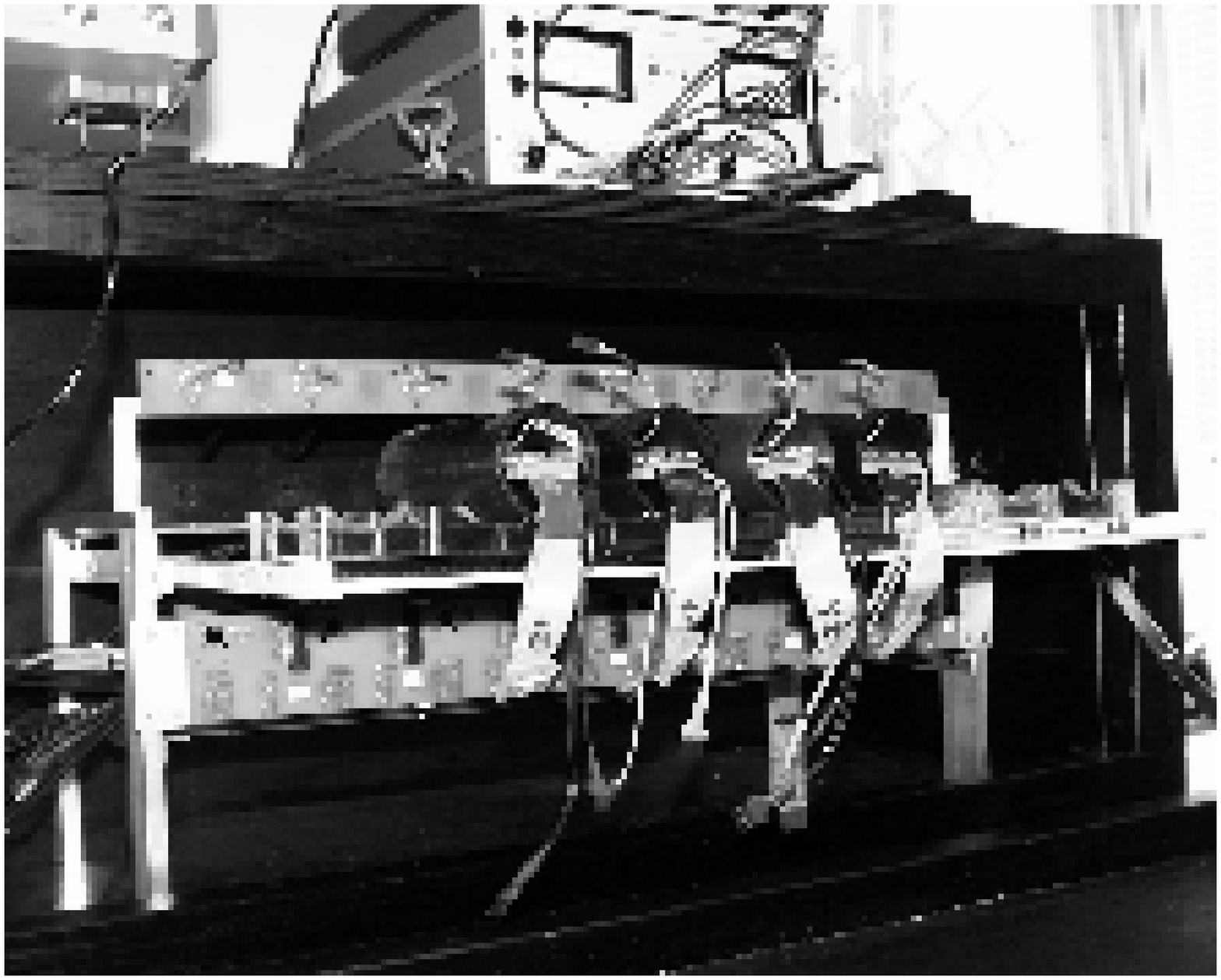,width=1.\linewidth,angle=0}
    \end{center}
  \end{minipage}
  \begin{minipage}{0.495\linewidth}
    \begin{center}
      \epsfig{file=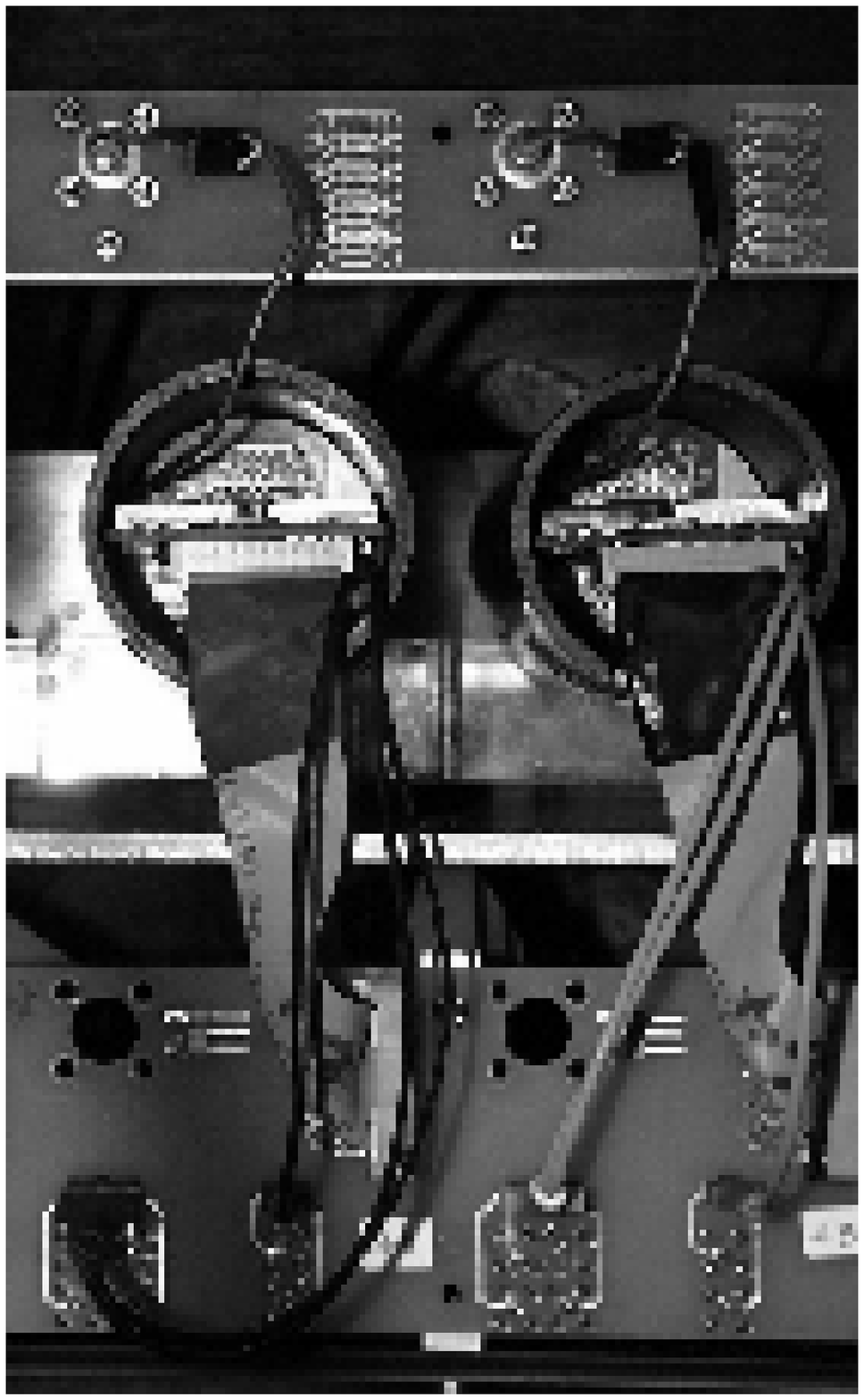,width=0.5\linewidth,angle=0}
    \end{center}
  \end{minipage}
  \caption{\sl\small The picture on the left shows the PMT-Blocks
    assembled inside the black box. The picture on the right shows a
    detail of the connection of the PMT-Blocks with two intermediate boards
    used for distributing signals, HV and controls.}
  \protect\label{fig:blackbox2}
  \protect\label{fig:specialboard}
\end{figure}

A schematic diagram of the test bench is shown in Figure \ref{fig:testbench}. 
Up to eight PMT-Blocks can be housed in special cradles inside a 
light-tight box (Figures \ref{fig:blackbox} and \ref{fig:blackbox2}). 
The Blocks are connected to the High Voltage distribution, 
to the signal and to the control cables, which are routed through the 
motherboard as they are in the detector.
In absence of a {\it drawer} 
the connection between the PMT-Blocks and the motherboard is provided by
special home-made boards (Figure \ref{fig:blackbox2}). 

The controls and readout proceed through the motherboard with a 
protocol which is the same as the one used in the detector. 
 
\begin{figure}[ht]
  \begin{center}
    \epsfig{file=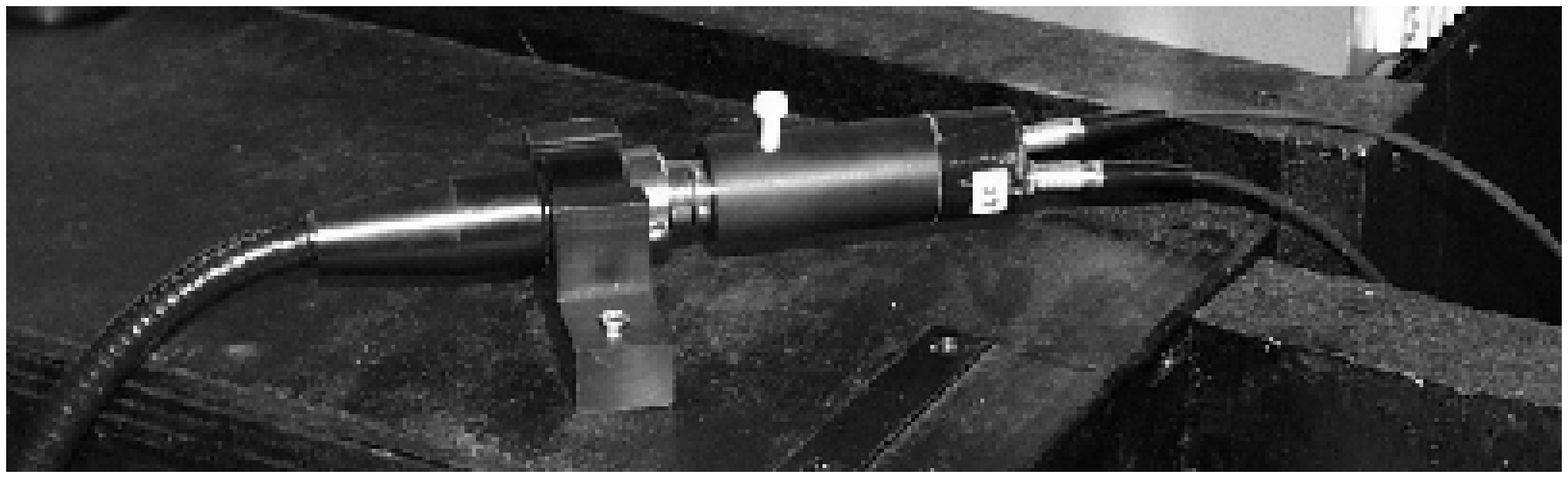,width=0.9\linewidth,angle=0}

    \epsfig{file=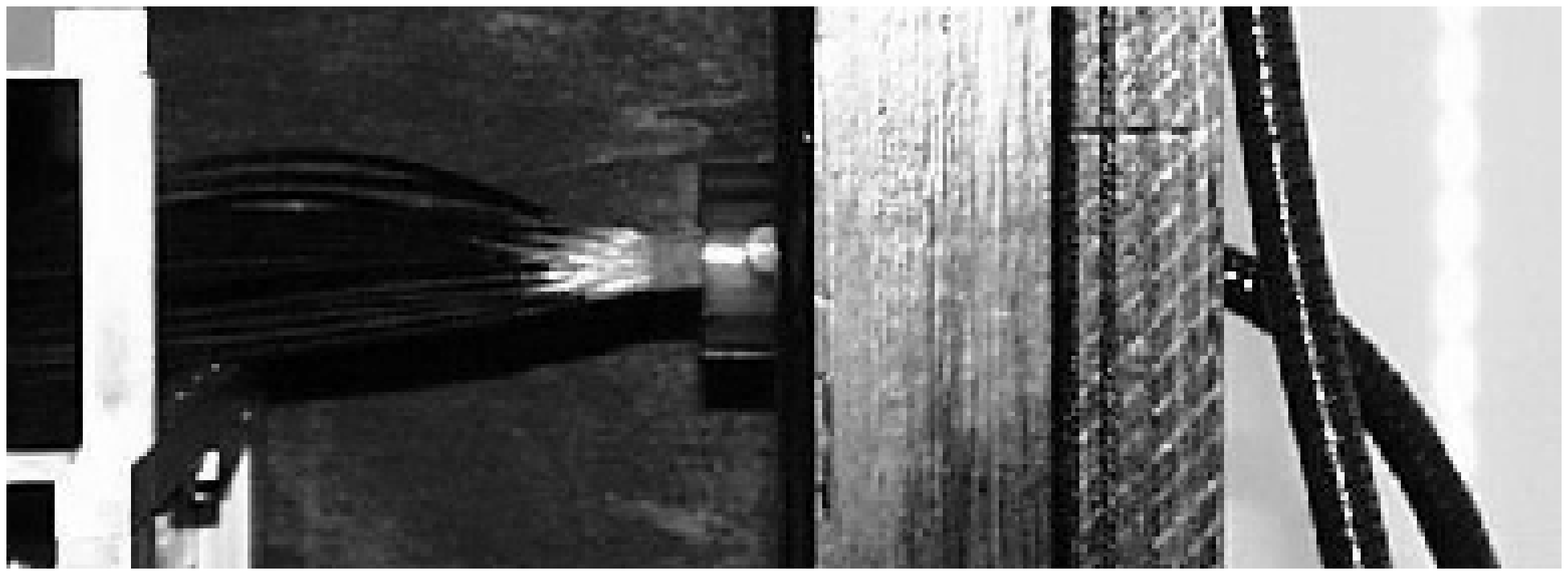,width=0.9\linewidth,angle=0}
    \end{center}
  \caption{\sl\small 
The picture on the top shows a detail of the light source
    system. Two LEDs, one operated in DC and the other in pulsed mode
    are assembled inside a black cylinder and coupled to a liquid fiber
    to take the light into the black box.
The picture on the bottom shows a detail of the
    coupling of the liquid fiber to the fiber bundle. The fiber bundle
    is made of 32 clear fibers which are distributed four in each PMT.} 
  \protect\label{fig:liquidbundle}
  \protect\label{fig:lighthouse}
\end{figure}

The PMTs are illuminated by two 
LEDs\footnote{\it Ledtronix PB280CWB1K}, one operating in continuous 
and the second in pulsed mode. 
The LEDs are housed in a black plastic cylinder and face a 
liquid fiber  
which is connected, inside the black box, 
to a fiber bundle, in order to distribute the light to each PMT 
(Figure \ref{fig:liquidbundle}).
The light source is very simple and was designed and built in
our labs. 
The amount of light each PMT receives has been roughly equalised 
among the eight PMTs using neutral optical filters. 

The DAQ of the test bench comprises the control and 
readout electronics  located in one NIM and one VME crate.
A single board PC 
(FIC 8232\footnote{\it CES, FIC8232 Fast Intelligent Controller}) 
located in the VME crate has been programmed to control the sequence of 
operations.
Control and pulse signals are produced by an I/O 
register\footnote{\it CAEN V262}.
The I/O bits of the register control the status of the HV power
supply\footnote{\it CAEN N472}.
The signal to pulse the
LED and to STOP the ADCs is generated acting on the I/O bits as well.

A VME Fermi ADC\footnote{\it CAEN V571} measures the light in the
pulsed mode (i.e. the {\it signal OUT} from the {\it 3-in-1} card of each PMT).
A second ADC\footnote{\it CAEN V265} is used to measure the PMT current when
operating in continuous light mode ({\it integrator OUT} from the {\it 3-in-1}
card).
Though the two ADC models are
not optimal for this task, they were available and could be easily 
included in the test bench.
They accomplished their assigned task in a satisfactory way.

A receiver card ({\it blue box} in Figure \ref{fig:testbench}) converts the
fast differential signals from the {\it 3-in-1} cards to single ended ones, 
and interfaces the {\it 3-in-1} to the Fermi ADC.

A second detector-specific unit 
({\it 3-in-1 controller} in Figure \ref{fig:testbench}) is used to set the
mode of operation of the {\it 3-in-1} cards and also to provide the 
charge injection
for the tests of the {\it 3-in-1} electronics.

As far as the LED drivers for pulse and continuous light are concerned,
we have preferred 
two home-made circuits to commercial units. 
A VME DAC unit sets the continuous light of the LED to the desired
intensity.
A NIM circuit generates fast (20-30 ns) pulses, whose amplitude can be
modulated by a DAC (0,+60 Volts) and is triggered by one bit of the I/O
register.

\begin{figure}[ht]
  \begin{center}
    \epsfig{file=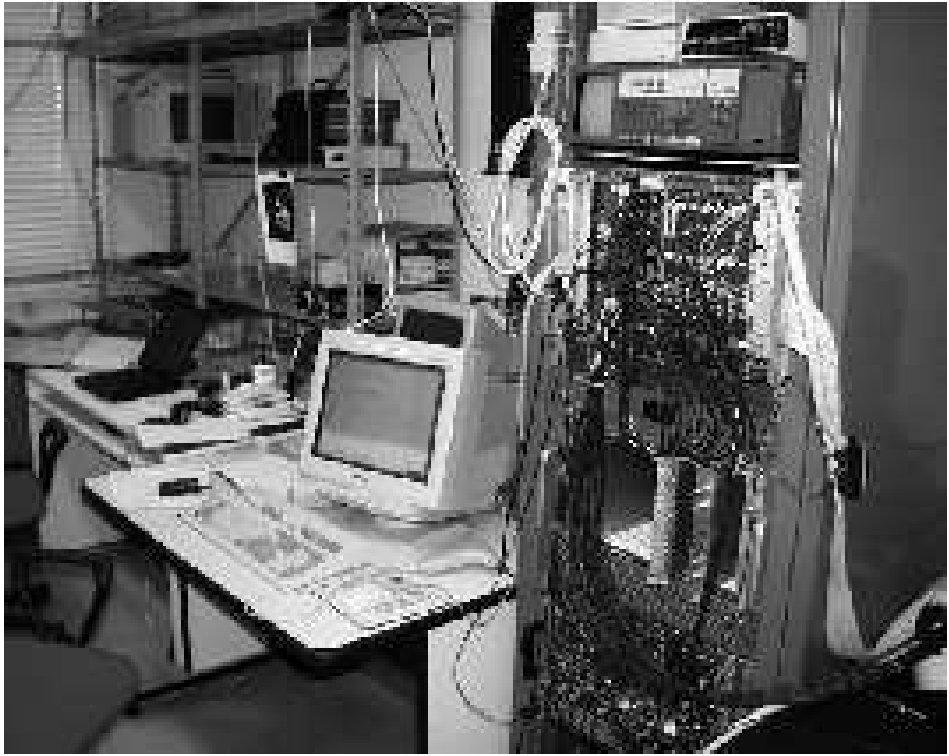,width=0.9\linewidth,angle=0}
  \end{center}
  \caption{\sl\small Overall view of the DAQ system.}
  \protect\label{fig:DAQ}
\end{figure}

The DAQ system was required to be simple and inexpensive, hence it had to be
designed with a few detector specific, home-made units.
The whole system is shown in Figure \ref{fig:DAQ}. 

\section{Protocol}
The sequence of operations needed for testing the PMT-Blocks
is described in this section.
As already mentioned,
the PMT-Block is an assembly of three main parts: a light mixer, a PMT
with its divider and the F/E electronics (the {\it 3-in-1} card). 

The {\it 3-in-1} cards form the front end of the electronics chain.  They
provide 3 basic functions: a fast pulse shaping of the PMT output with two
gains to provide the dynamic range, a
slow integrator circuit for measuring DC current for detector monitoring
\cite{bib:TDR},
and a charge injection system for electronics calibration. 
These three functions are controlled by an on-board logic driven 
by a remote controller. 

The test starts checking the functionality of the {\it 3-in-1} card,
independently of the PMT, by injecting a precise quantity of charge into the
{\it 3-in-1} input and recording the amplified and shaped output pulses. 
Then, short light pulses are sent to the PMT to check its 
performances.
Finally,
continuous light of different intensities is used to test the Integrator in
different operating conditions. 

\subsection{The Test of the Charge Injection System}
During the Charge Injection System test the high voltage and light
system are OFF and the {\it 3-in-1} card is set in the charge injection
configuration. Different charges are then injected, covering the dynamic range
of the device, both in the {\it low} and the {\it high} gain. Output signal 
from the low gain, high gain and trigger channels are recorded. 
At the end of this test, a linear fit of the measured output signal as function
of the injected charge,
is performed.
The low and high gain of each channel are the slope of the fit line and are 
stored on disk for further analysis.

\subsection{The test with light pulse}
During this part of the test the high voltage is ON and the light system
operates in pulsed mode. The aim is to simulate the working conditions
of the PMT in the experiment by flashing the PMT with light of different
intensities. 
In tests performed previously, all the PMTs have been characterized and
their nominal voltage has been defined (nominal HV is that for which the gain
of a given PMT is $10^5$).
However, for this test a common value of the HV (800 V) 
was preferred for simplicity.
Three different light levels were used to simulate: a) the signals produced in
the calorimeter by a minimum ionizing particle, b) a signal falling in
the energy region where the two gains overlap, and c) a jet with energy 
of about one TeV. 
For each PMT-Block, 5000 events are then recorded at each light setting. 
The pulse shape is recorded by the V571 flash ADC. The program performs the 
pulse measurement and the pedestal subtraction on-line. 
Two histograms for each PMT are
filled: the first one for the low  gain, the second one for the high gain. The
average value of each histogram and its r.m.s. are stored to disk.

\subsection{The Test of the slow integrator}
During the Slow Integrator test the high voltage is ON, the light
system is operated in DC mode and the 3-in-1 card is set to integrator mode. 
This operation will be used to monitor the calorimeter gain using {\it minimum
bias events} when LHC is running and the light produced by a $^{137}$Cs
source that, in special runs, will move through all of the calorimeter
cells \cite{bib:cesio}.
Three different levels of DC light are
used to simulate the pile-up events, the Cs source and a level of light 
in between. 
The average output current for each level (pedestal subtracted) 
is measured and stored on disk.

\subsection{Data analysis and decision}
The data analysis is performed on-line, right after the data
taking and, at the end of the test, the 
relevant information to the
PMT-Block is available in an ASCII file.
Since the aim of this test is to check the functionality of the PMT-Blocks
without performing their full characterization, the measured values
are checked to be within an acceptance range.
If this is the case, the PMT-Block is accepted; otherwise it is rejected 
and sent back for repair. 
A screen message indicates to the operator which
PMT-Blocks are to be rejected. The acceptance ranges used are summarized
in Table \ref{tab:range}.
\begin{table}[htb]
  \begin{center}
    \begin{tabular}{|l|c|c|c|}
    \hline
               & \multicolumn{3}{c|}{Acceptance window} \\
    \hline
    \hline
    CIS (gain ratio)  & \multicolumn{3}{c|}
{ $ 44 \leq G_{high}/G_{low} \leq 84 $} \\
    Fast Pulse:       & \multicolumn{3}{c|}{\(0 \leq r.m.s/mean \leq 0.25\)} \\
\hline
 DC Light        & Min Bias & Medium & $^{137}$Cs source \\ 
 Integrator      & $20 \leq Q \leq 400$ 
&  $50 \leq Q \leq 800$ 
&  $75 \leq Q \leq 1000$ 
\\
    \hline
\end{tabular}
\end{center}
\caption{\sl\small The acceptance range used for the three different tests.}
\protect\label{tab:range}
\end{table}
At the end of each measurement the ASCII file is imported in Microsoft
EXCEL\textsuperscript{\texttrademark}
which performs the statistics and the database functions.

\subsection{Sequence of operations}
The sequence of operations required to perform the tests
of a PMT-Block batch described above, is the
following:
\begin{enumerate}
\item an ensemble of eight PMT-Blocks are mounted into the black box (manual);
\item the cables of the PMT-Blocks are connected to the patch panel (manual);
\item the identification information of the PMT-Block are collected and
the database initialised (manual);
\item the CIS test is then performed and the results are stored to disk 
(automatic);
\item the fast pulse test is performed and the results stored to disk
  (automatic);
\item the slow integrator test is performed and the results stored to disk
  (automatic);
\item the results are analysed and the decision is made 
to accept or reject the PMT-Block
  (automatic);
\item finally the PMT-Blocks are removed from the test bench. The PMT-Blocks 
to be rejected are put aside 
  (manual).
\end{enumerate}
A complete test of a batch of 8 PMT-Blocks takes about 25 minutes from
the PMT insertion to their removal.

\section{Software}

The guidelines followed in the development of this test bench have
been stated in the introduction. The software has to fulfill all these
requirements, i.e. it has to be affordable, easy to use, stable,
portable and versatile.

The software was developed by following a layered approach. 
The operations to be performed on the VME data bus are two: 
write to and read from the bus. Every board is interfaced
by a small library which supplies the basic set of routines to operate
the board itself.

The layered structure of our software is the following:

\begin{description}
\item[--] {\sl kernel} -- This is the part of the code that interfaces
  to the menu, data and file handling and deals with error
  management. It takes care of the VME bus initialization and
  read/write functions. This is the only platform dependent part of
  the code and must be written accordingly to the hardware used to
  interface the VME bus;
\item[--] {\sl libraries} -- Each VME board has its own library of
  functions that are responsible for initialization, control and data
  acquisition. If a new board is added, a new library has to be
  written. The debugging is easy as only the basic functions have to
  be checked, while more complex functions can be obtained by suitable
  scripts. At this level, the software is platform independent and can
  be moved from one system to another without changing the code if the
  system supplies an ANSI-C compliant compiler;
\item[--] {\sl scripts} -- This is the part dealing with the
  implementation of a specific test. The user takes advantage of the
  hardware functions without worrying how operations are performed and
  has only to concentrate on the measurements. This part is, of
  course, platform independent.
\end{description}

The software has been originally developed on a Cetia PowerPC board (CVME 604)
with LynxOS~2.4 operating system. 

Every hardware board is activated by a set of commands available from
a sub-menu in the DAQ program. These routines are used for simple
operations like setting the number of boards present in the crate, setting
the VME address to access them and the basic read/write functions.

The DAQ system is essentially a command line driven program, with
simple operations available from menus and sub-menus that perform
specific actions. Multiple commands can be written on the same line
and the program takes care of splitting them into single tokens. 
These commands can be grouped in macros that the program is
able to execute. Macros can be called and nested and complicated
sequences of commands are easily obtained, thus providing a very high
level of flexibility. Any kind of test bench can be implemented
without too much trouble, provided the right macros are written.

To handle the data coming from the VME boards, we have included in the
system a histogram handling package \`a la HBOOK~\cite{bib:hbook} and
a data handling package \`a la SIGMA~\cite{bib:paw}, to perform
simple mathematical operations as well as file handling. 
A preliminary analysis can be performed on-line but data can be
written on disk for an accurate, off-line, analysis.


During the years this software has been tested on several systems, 
from Motorola 68x to PowerPC and Intel based platforms. 
Besides the usage in the ATLAS experiment, the same software has been 
used in several occasions; from data acquisition in the test beam setup 
for the AMS electromagnetic calorimeter~\cite{bib:ams-ecal} 
to the trigger interface in the framework of the 
MAGIC telescope~\cite{bib:magic-l2}.

An example of the scripts used in the test is shown in the Appendix.


\section{Results from the PMT Block test}

In the following some results from the test of 
about 10000 PMT-Blocks are presented.
\subsection{Charge Injection System Test}
\begin{figure}[ht]
  \begin{center}
    \epsfig{file=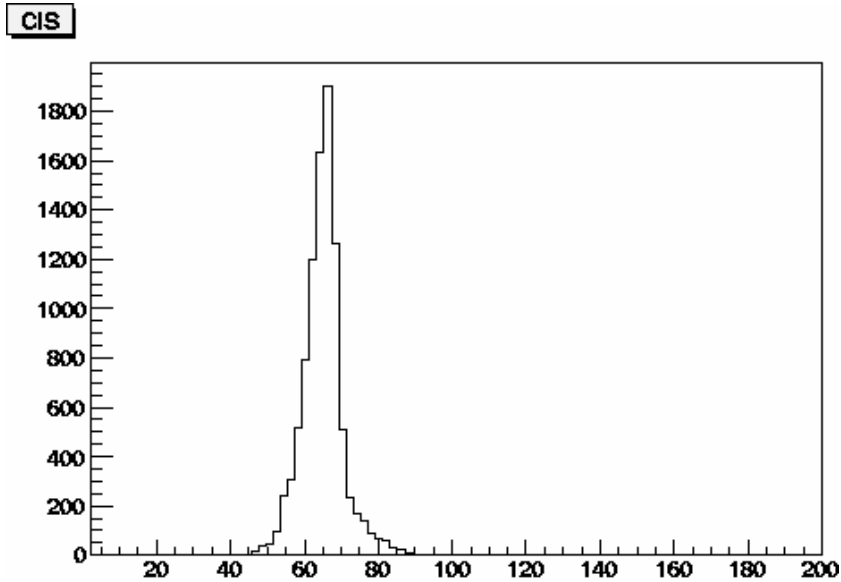,width=0.55\linewidth,angle=-90}
  \end{center}
  \caption{\sl\small 
Distribution of the CIS slope. 
The acceptance window is the interval $44-84$. 
}
\protect\label{fig:cisresults}
\end{figure}

In the charge injection test, the important parameter 
is the ratio ($R$) of the gains measured in the 
high gain mode and low gain mode of the {\it 3-in-1} 
card. 
The distribution of $R$ is shown
in Figure \ref{fig:cisresults}. 
The mean value is 65.19, very close to the 
nominal value of 64. 
It should be noted that the resolution of this measurement does not reflect
the quality of the hardware \cite{bib:treinuno},\cite{bib:interface}, it
rather represents the accuracy of this test procedure.
 \subsection{Light Pulse test} 
\begin{figure}[ht]
  \begin{center}
    \epsfig{file=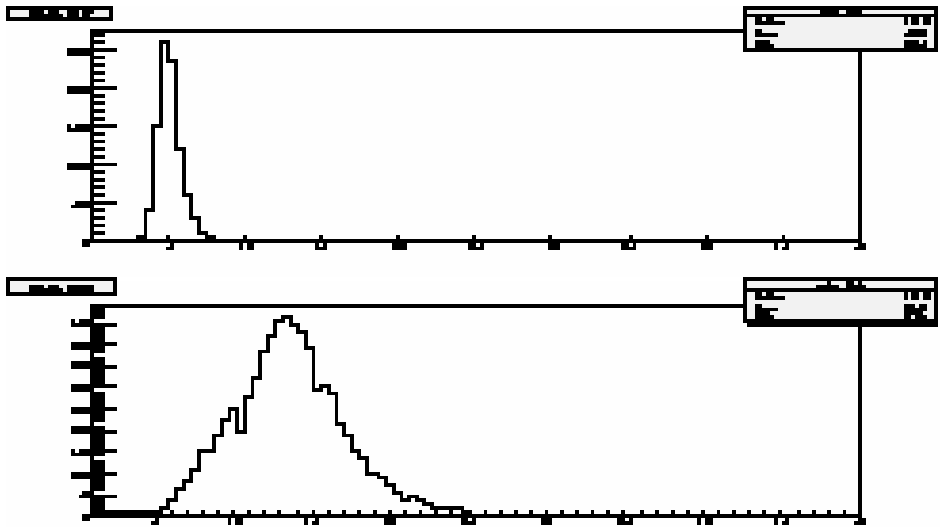,width=0.5\linewidth,angle=-90}
  \end{center}
  \caption{\sl\small Top: Distribution of $rms/mean\cdot 100$ for low gain 
working conditions. Bottom: Distribution of the $rms/mean\cdot 100$ for high
gain working conditions. 
}
\protect\label{fig:pulseresults}
\end{figure}

The output of the Pulse Test provides the mean value and the r.m.s. 
for the high and low gain mode of the {\it 3-in-1} card. 
For a PMT-Block to pass this part of the test, the value of the ratio:
r.m.s./mean has to be less than 25\%.
In Figures \ref{fig:pulseresults} $a)$ and $b)$ the distributions 
of this quantity for low and 
high gain respectively are shown.
The effective resolution shown in the plots is due in part to the
photo-statistics and in part to the different share of light
in the different channels. 
It is clear that in both cases this condition is fulfilled for 
the vast majority of the PMT-Blocks. 
\subsection{Integrator test} 
\begin{figure}[ht]
  \begin{center}
    \epsfig{file=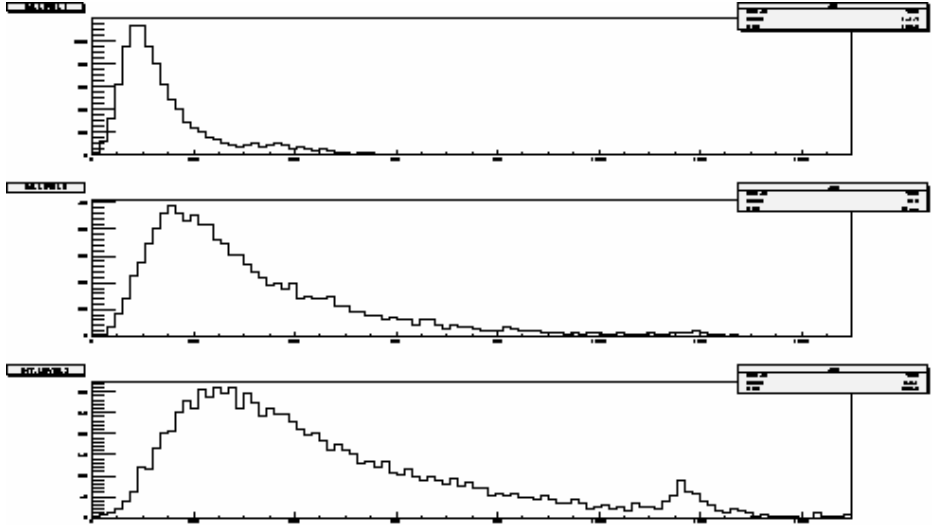,width=0.5\linewidth,angle=-90}
  \end{center}
  \caption{\sl\small 
Distribution of the mean value of INT1 (a), INT2 (b) and  INT3 (c). 
The acceptance window, for each plot, is the interval 
$(20,400)$, $(50,800)$ and $(75,1000)$.}
  \protect\label{fig:pulseresults1}
\end{figure}

\begin{figure}[ht]
  \begin{center}
    \epsfig{file=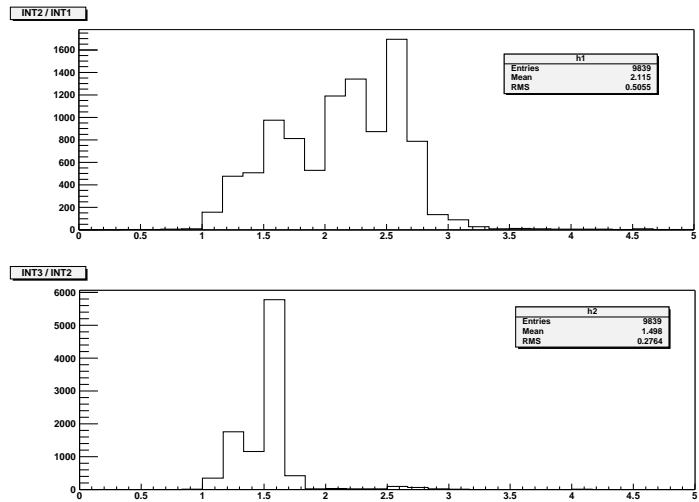,width=0.75\linewidth,angle=0}
  \end{center}
  \caption{\sl\small 
Distribution of the ratios of  INT2/INT1 (upper plot), 
and INT3/INT2 (lower plot)}
  \protect\label{fig:ratios}
\end{figure}

In this third test, three different DC light levels are used. 
For each level, the current is measured and 
its mean value is recorded together with
the corresponding r.m.s..
The aim of the procedure is to observe an 
increase of the mean value as a consequence of the increase of 
the light level. 
The mean values of the three levels are denoted as INT1, INT2 and INT3 
from the lower to the higher, respectively. 
The distribution of the mean values of the
three quantities is shown in
Figure \ref{fig:pulseresults1}. 
The acceptance ranges of the three mean values were given in
Table \ref{tab:range}. 
The distributions of figure \ref{fig:pulseresults1} are quite wide
because no attempt 
was done to
equalize the gain of the PMTs. It was verified that the system is 
behaving 
correctly by
plotting the ratios of the integral event by event, as is shown in
Figure \ref{fig:ratios}. 

\subsection{Fraction of rejected PMT-Blocks} 
The fraction of the PMT-Blocks rejected by each of 
the three tests in the first pass is shown in 
Table \ref{tab:rejectedblocks}; the same table also summarizes 
the percentage of rejected PMT Blocks. 

\begin{table}[htb]
  \begin{center}
    \begin{tabular}{|l|c|}
    \hline
       Test  & Fraction of rejected PMT-Blocks (\%) \\
    \hline
    \hline
    CIS slope   &  2.1\\ \hline
    Fast Pulse test & 0.16 \\ \hline
    Integrator Test  &  0.16 \\ \hline
    Total Fraction rejected & 2.42 \\ \hline \hline
    \end{tabular}
   \end{center}
  \caption[dai]{\sl\small 
Fraction of rejected PMT-Blocks in the first pass
}
  \protect\label{tab:rejectedblocks}
\end{table}

\section{Conclusions}
 
A test bench for the PMT-Blocks employed in the 
ATLAS Tile Calorimeter has been developed. 
The aim of these tests was to ensure the functionality of the PMT-Blocks 
before their final insertion in the super-drawers. 
The test allows fast and accurate enough measurements 
of the main properties of a 
complete PMT-Block (light mixer / PMT and divider / {\it 3-in-1} card).  
The ability of performing different tests, each of them checking a 
special operation mode of the PMT-Block, is another important 
characteristic of the system.
This test bench has been used for the quality control of 10300 PMT-Blocks 
that will be used in the ATLAS experiment.  
The different operational modes of the Blocks were checked by 
three different tests, in order to ensure their full functionality. 
The DAQ system consists of basic routines, each one 
corresponding to a different electronic instrument housed in the 
VME crate, and it is based on a structure which permits high level calls 
from a non-compiled code. 
The fraction of the rejected PMT-Blocks in the first pass was about 2.5\%. 

\section{Acknowledgments}
We want to thank the Tile Cal group of the University of Chicago, 
in particular K. Anderson, for
the help and advice in the operations of the {\it 3-in-1} cards and for
lending us the $3in1$ controller and its software.
Our thanks also to Tile Cal group of Stockholm that lend us the {\it 3-in-1}
receivers. All the Tile Cal collaboration has encouraged this effort and we
want to thank all those that have helped us with advice and support in
understanding the details of the PMT-block system.
    
This work has been partly supported by the Italian Istituto Nazionale di
Fisica Nucleare (INFN) and the University of Pisa, 
by the Greek General Secretariat 
for Research and Technology (GGSRT), by EEC RTN contract 
HPRN-CT-00292-2002, 
by INTAS-CERN grant N.O. 03-52-6477,
by the Special Account for 
Research Grants of the National and Kapodistrian University of Athens 
and by the EPEAEKII (Operational Program for Education and 
Initial Vocational Training) program in the framework of the projects 
Pythagoras and Iraklitos. 

\appendix
\section{The DAQ software}
In this section we show some examples of how the scripts
used in the Data Acquisition for these tests look like.

The scripts are ordered in a hierarchy. The top level scripts 
organize the sequence of the different tests.
The first script is the {\it test.uic}:

{\center \scriptsize \begin{verbatim}
!!!!!!!!!!!!!!!!!!!!!!!!!!!!!!!!!!!!!!!!!!!!!!!!!!!!!!!!!
!   Macro Name: test.uic				!
!!!!!!!!!!!!!!!!!!!!!!!!!!!!!!!!!!!!!!!!!!!!!!!!!!!!!!!!!
!   CALLED BY: user input 				!
!   CALLS: 	test_cis.uic				!
!		test_pulse.uic				!
!		test_int.uic				!
!		test_merge.uic				!
!!!!!!!!!!!!!!!!!!!!!!!!!!!!!!!!!!!!!!!!!!!!!!!!!!!!!!!!!
!							!
!   This is the macro starting the PMT block test	!
!   							!
!   Run this one with the command @test.uic in vmetest	! 
!   to perform a complete test.				!
!   Every newly implemented test part should be called 	!
!   from here.						!
!							!
!!!!!!!!!!!!!!!!!!!!!!!!!!!!!!!!!!!!!!!!!!!!!!!!!!!!!!!!!
message "************************************************\n"
message "*   The PMT Block Test Bench                   *\n"
message "*   Version 1.0          May 30th 2000         *\n"
message "************************************************\n"
!                        CIS Test
@test_cis.uic
!                        Pulsed Test
@test_pulse.uic
!                        Integrator Test
@test_int.uic
!                        Summarize results on test.dat
@test_merge.uic
message "Test is Finished\n"
message "RENAME the output file\n"
\end{verbatim}
\newpage

}

The script consists of several calls to other scripts. The first is the
script {\it test\_cis.uic} which is listed below:

{\scriptsize \begin{verbatim}
!!!!!!!!!!!!!!!!!!!!!!!!!!!!!!!!!!!!!!!!!!!!!!!!!!!!!!!!!
!   Macro Name: test_cis.uic                            !
!!!!!!!!!!!!!!!!!!!!!!!!!!!!!!!!!!!!!!!!!!!!!!!!!!!!!!!!!
!   CALLED BY: test.uic	                                !
!   CALLS:  cis_fast.uic     				!
!!!!!!!!!!!!!!!!!!!!!!!!!!!!!!!!!!!!!!!!!!!!!!!!!!!!!!!!!
!                                                       !
!   This is the macro for Charge Injection calibration  !
!							!
!   The real work is done in cis_fast.uic		!
!   This macro is only the final interface. 		!	
!							!
!   call @test_cis.uic from vmetest to perform only 	!	
!   CIS test (for debugging purposes)			!
!							!
!!!!!!!!!!!!!!!!!!!!!!!!!!!!!!!!!!!!!!!!!!!!!!!!!!!!!!!!!

message "*************************\n"
message "* CIS calibrations      *\n"
message "*************************\n"
!                     Call to cis_fast.uic
@cis_fast.uic
!                     Print the results obtained for online monitoring
message @gain_low1.dat	message "\n"
message @gain_low2.dat	message "\n"
message @gain_low3.dat	message "\n"
message @gain_low4.dat	message "\n"
message @gain_low5.dat	message "\n"
message @gain_low6.dat	message "\n"
message @gain_low7.dat	message "\n"
message @gain_low8.dat	message "\n"
message @gain_high1.dat	message "\n"
message @gain_high2.dat	message "\n"
message @gain_high3.dat	message "\n"
message @gain_high4.dat	message "\n"
message @gain_high5.dat	message "\n"
message @gain_high6.dat	message "\n"
message @gain_high7.dat	message "\n"
message @gain_high8.dat	message "\n"

\end{verbatim}
\newpage

}

which, in turns calls {\it cis\_fast.uic}:

{\center \scriptsize \begin{verbatim}
!!!!!!!!!!!!!!!!!!!!!!!!!!!!!!!!!!!!!!!!!!!!!!!!!!!!!!!!!
!   Macro Name: cis_fast.uic                            !
!!!!!!!!!!!!!!!!!!!!!!!!!!!!!!!!!!!!!!!!!!!!!!!!!!!!!!!!!
!   CALLED BY: test_cis.uic                             !
!   CALLS:	cis_hist_book.uic                       !
!           	cis_init2.uic				!
!		cis_hist_reset.uic   			!	
!		cis_event_low.uic			!
!		cis_hist_fill_low.uic			!
!		cis_stack.uic 				!
!		cis_init1.uic				!
!		cis_event_high.uic			!
!		cis_hist_fill_high.uic			!
!   VARIABLES: X0 --> injected charge                   !
!!!!!!!!!!!!!!!!!!!!!!!!!!!!!!!!!!!!!!!!!!!!!!!!!!!!!!!!!
!                                                       !
!   This is the macro for Charge Injection calibration  !
!   Version using the fast Fermi reading to save cpu    !
!   time. 						!
!							!
!   COMMENT: The fast fermi reading returns as a result !
!   the peak value subctracted by the average read      !
!   of the first 5 fermi samplings.			!	
!							!
!   The real cis work is done here and divided into     !
!   submacros each doing simple actions.		!
!!!!!!!!!!!!!!!!!!!!!!!!!!!!!!!!!!!!!!!!!!!!!!!!!!!!!!!!!
!
!                                Book histos used for the cis test
@cis_hist_book.uic
!                                Inizialize 3 in 1 controller 
!                                Each of the 8 tubes are initialized 
!                                by a call to cis_init2.uic
!
3in1
        @cis_init2.uic
        zone 0 sector 0 tube 47 address
!                                Repeat for all tubes (not listed)
return
!                                Reset and inizialize Fermi modules. 
Fermi
	select 1 reset sample_write 30 delay_write 0
!                                Repeat for all Fermis (not listed)
return
!
!                                Low gain CIS calibration: 
!                                Injected charged is 0, 200, 400, 600, 800
message 0\n
wait 1
sigma <   0 > X0 ret
!                                Acquire 10 CIS events. 
loop 10 @cis_event_low.uic
!                                The mean is stored in another histo
!                                (1111 for pmt 1) as a function of Qinj. 
!
@cis_hist_fill_low.uic
!                                Repeat for all  Change injected values
!                                (not listed)
!                                Put results into files. 
sigma
	fopen gain_low1.dat
!                                write tube number
        fwrite "1\t"
!                                Convert the value to appropiate units
        < 1.249 > x6 pull
!                                Linear fit to the histo 
	fit 1111 @cis_stack.uic fclose
!                                Repeat for the other tubes 
!                                (same but not listed)
return
!                                High gain calibration (same but not listed).

\end{verbatim}

}

This last script consists of several calls to ancillary scripts to
manage histograms. The most relevant call is the one to {\it
FERMI}. This routine performs basic I/O operation on the VME bus. This
routine is coded in C and, part of it is listed below:

{\center \scriptsize \begin{verbatim}

/* 
 *  This command performs a read to the Fermi V571 samples
 *  and returns the difference between the maximum sampling value 
 *  and the mean of the first 5 samplings
 *
 *  This is a replacement of the READ command to speed up the test
 *
 */

if(strcmp(verb,"FASTREAD")==0) /* User command is FASTREAD */
{
/*
 *  Get the qualifiers associated with the FASTREAD command
 *  Debug  : 1 Debug is on / 0 Debug is off
 *  Channel: number of channel to be read (1,2,3)
 */  

  if((status = get_qualifier("PMT_Fermi_Menu", "FASTREAD", "CHANNEL",&nch))==0){}
  if((status = get_qualifier("PMT_Fermi_Menu", "FASTREAD", "DEBUG",  &deb))==0){}
/* 
 * arr[256] is used to store the Fermi samplings information. It is first 
 * set to zero and then filled up by a call to Fermi_wait_for_trigger
 */

  uzero(arr,256);
  err=Fermi_wait_for_trigger( nch, arr);
/* 
 * Detect eventual errors
 */
  if(err>0)
  {
          error_logger( "Fermi_wait_for_trigger", ERWARN, "Timeout error");
          return;
  }
  else
  {
/* 
 * Debug if it is requested
 */
     if(deb>0) printf("Fermi %ld Channel=%ld\n", this_fermi, nch);
/*
 * average is the average of the first 5 samplings used as a pedestal
 */
     average=(float)arr[0]+arr[1]+arr[2]+arr[3]+arr[4];
     average /= 5.;
/*
 * Look for the maximum voltage in the 256 Fermi samplings
 */
     vmax= 0.;
     for(i=0;i<sample;i++)
     {
        if(deb>0)printf("%d %d\n",i,*(arr+i));
        if((float)arr[i]>vmax)vmax=(float)*(arr+i);
     }
/* 
 * Push into the BBmenu stack the relevant information, i.e. the
 * difference between the maximum and the average of the first 5 samplings
 */
     pushStack(vmax-average);
  }
}

\end{verbatim}

}

\end{document}